\documentclass[aps,pra,reprint,superscriptaddress,onecolumn]{revtex4}

\usepackage{graphicx}
\usepackage{bm}
\usepackage{bbold}

\usepackage{color}

\begin{document}

\title{Incomplete spontaneous decay in a waveguide caused by polarization selection}
\author{A. S. Kuraptsev}
\email[]{aleksej-kurapcev@yandex.ru}
\affiliation{\small Peter the Great St. Petersburg Polytechnic University, 195251, St. Petersburg, Russia}
\author{I. M. Sokolov}
\email[]{ims@is12093.spb.edu}
\affiliation{\small Peter the Great St. Petersburg Polytechnic University, 195251, St. Petersburg, Russia}

\date{\today}

\sloppy



\begin{abstract}
Spontaneous decay of an excited atom in a waveguide is essentially modified by the spatial structure of vacuum reservoir. This is particularly exciting in
view of a range of applications for quantum information science. We found out that spontaneous decay can be incomplete, so the time dependence of the excited state population asymptotically approaches to a nonzero value, under the conditions when the atomic transition frequency is larger than the cutoff frequency of a waveguide and far from the vicinities of the cut-offs. Discovered effect is explained by the emergence of the dark state, which is non-decaying due to polarization selection rules. It was revealed for single-mode waveguide with rectangular cross section both in single-atom case and diatomic case when the long-range dipole-dipole interaction plays a significant role.

\end{abstract}


\maketitle
\section{Introduction}
Atomic spontaneous decay is one of the most fundamental phenomena of quantum electrodynamics, taking place due to the coupling between an atom and vacuum reservoir. Now it is well understood that changing the properties of the reservoir one can affect the spontaneous decay process. Thus, placing an atom in a cavity or waveguide leads to significant alteration of the decay rate \cite{1,2,Kleppner,Klimov1,Klimov2,Yudson,Kupr1,Kupr2}. In the case of a waveguide, this effect dramatically depends on the ratio between the resonant frequency of atomic transition $\omega_0$
and the cutoff frequency of a waveguide $\omega^{c}$. In particular, in the well-known paper of Kleppner \cite{Kleppner}, strong suppression of spontaneous decay under condition $\omega_0<\omega^{c}$ was predicted. The same effect takes place for an excited atom in photonic band gap crystals when $\omega_0$ falls in the range of the frequency gap of the environment \cite{Bykov,Lambropoulos}. When $\omega_0$ approaches the vicinities of a photonic crystal bandgap where the local density of states is not smooth, the suppression of decay is
not complete, but only partial, so only some part of the energy of initial atomic excitation is transferred to the electromagnetic field and other part remains in the atomic system. This effect is explained by frequency selection and can be described in the framework of two-level atom formalism neglecting the vectorial nature of electromagnetic waves, as it was done in Ref. \cite{Lambropoulos}.

In this paper we report the discovery of incomplete spontaneous decay in a waveguide when the atomic transition frequency $\omega_0$ is larger than $\omega^{c}$ and far from the vicinities of the cut-offs of the waveguide modes. We explain the revealed effect by polarization selection.
Our results raise the issue of the role that polarization plays in the problem of spontaneous decay in structured reservoirs. Discovered effect might be of general interest in different topics of quantum physics, in particular, quantum information storing and processing \cite{waveduide_group3_2, waveduide_group3_1}. It can be useful for the development and improvement of many quantum devices based on atomic systems in a waveguide, such as single-photon switching \cite{waveduide_group3_3,waveduide_group3_4,waveduide_group3_5}, routers \cite{waveduide_group3_6}, transistors \cite{waveduide_group3_7,waveduide_group3_8,waveduide_group3_9}, frequency comb generators \cite{waveduide_group3_10}, and single-photon frequency converters \cite{waveduide_group3_11}.

\section{Basic assumptions and approach}
The theory employed here considers an ensemble of point-like motionless atoms in a waveguide. This model is excellent for ensembles of impurity atoms embedded in a transparent dielectric under low temperatures that, therefore, provide a fantastic and practically
realizable playground for testing the theory \cite{Naumov1,Naumov2}. For concreteness, we assume
that the atoms are equal, having a nondegenerate ground state $|g_{i}\rangle$ with energy $E_{g}$ and the total angular momentum $J_{g}=0$ and an excited state $|e_{i}\rangle$ with $E_{e}=E_{g}+\hbar\omega_{0}$, $J_{e}=1$ and natural free space linewidth $\gamma_{0}$ ($\hbar$ is the Planck's constant and the index $i=1,...,N$
denotes quantities corresponding to the atom $i$ among $N$
atoms). The excited state is thus triply degenerate and splits in three Zeeman sublevels $|e_{i,m_{J}}\rangle$, which differ by the angular momentum projection on the quantization axis $z$ -- $m_{J}=-1,0,1$. For convenience, let us choose $z$ axis coinciding with the axis of a waveguide. Assuming the walls of a waveguide to be perfectly conductive (i.e. neglecting the absorption), we can write the non-steady-state Schrodinger equation for the wave function of the joint system, which consists of the atoms and the electromagnetic field in a waveguide, including vacuum reservoir. This system is described by the following Hamiltonian \cite{Cohen}:
\begin{eqnarray}\label{lett1}
  \widehat{H}&=&\sum_{i=1}^{N}\sum_{m_{J}=-1}^{1}\hbar\omega_{0}|e_{i,m_{J}}\rangle\langle e_{i,m_{J}}| \nonumber \\
  &+&\sum_{\textbf{k},\alpha}\hbar\omega_{k}\left(\widehat{a}_{\textbf{k},\alpha}^{\dagger}\widehat{a}_{\textbf{k},\alpha}+\frac{1}{2}\right)-\sum_{i=1}^{N}\widehat{\textbf{d}}_{i}
  \cdot\widehat{\textbf{E}}\left(\textbf{r}_{i}\right) \nonumber \\
  &+&\frac{1}{2\epsilon_{0}}\sum_{i\neq j}^{N}\widehat{\textbf{d}}_{i}\cdot\widehat{\textbf{d}}_{j}\delta\left(\textbf{r}_{i}-\textbf{r}_{j}\right),
\end{eqnarray}

where the first two terms correspond to noninteracting
atoms and the electromagnetic field in an empty waveguide, respectively, the
third term describes the interaction between the atoms and
the field in the dipole approximation, and the last, contact
term ensures the correct description of the electromagnetic
field radiated by the atoms \cite{Cohen}. In Eq. (\ref{lett1}), $\widehat{a}_{\textbf{k},\alpha}^{\dagger}$ and $\widehat{a}_{\textbf{k},\alpha}$ are the
operators of creation and annihilation of a photon in the corresponding mode, $\omega_{k}$ is the photon frequency, $\widehat{\textbf{d}}_{i}$ is the dipole operator of the atom $i$, $\widehat{\textbf{E}}\left(\textbf{r}\right)$ is the electric displacement vector in a waveguide, and $\textbf{r}_{i}$ is the position of the atom $i$.

Field operator $\widehat{\textbf{E}}(\textbf{r})$ can be obtained on the basis of well known classical mode expansion of the electromagnetic field in a waveguide \cite{Jackson} followed by standard quantization \cite{Raudorf}. The specific form of this operator is determined by the cross section of a waveguide. For concreteness, we assume the rectangular cross section with sizes $a$ and $b$. In this case, $\widehat{\textbf{E}}(\textbf{r})$ is given as follows:
\begin{eqnarray}\label{4}
  \widehat{\textbf{E}}(\textbf{r})&=&\sum_{\textbf{k},\alpha}\sqrt{\frac{\hbar}{2\omega_{k}}}\textbf{E}_{\textbf{k},\alpha}\left(x,y\right) \nonumber \\
  &\times&\exp\left(\texttt{i}k_{z}z\right)\widehat{a}_{\textbf{k},\alpha}+\text{H.c.},
\end{eqnarray}
where $\alpha$ denotes the type of waveguide mode -- TE (transverse electric) or TM (transverse magnetic), $\texttt{i}$ means imaginary unit.
\begin{eqnarray}\label{5}
  E_{\textbf{k},TE}^{x}\left(x,y\right)&=&-\frac{\texttt{i}k_{n}k}{k_{m}^{2}+k_{n}^{2}} \nonumber \\
  &\times& B_{mn}
  \cos{\left(k_{m}x\right)}\sin{\left(k_{n}y\right)},
\end{eqnarray}
\begin{eqnarray}\label{6}
  E_{\textbf{k},TE}^{y}\left(x,y\right)&=&\frac{\texttt{i}k_{m}k}{k_{m}^{2}+k_{n}^{2}} \nonumber \\
  &\times& B_{mn}\sin{\left(k_{m}x\right)}\cos{\left(k_{n}y\right)},
\end{eqnarray}
\begin{equation}\label{7}
  E_{\textbf{k},TE}^{z}\left(x,y\right)\equiv0,
\end{equation}
\begin{eqnarray}\label{8}
  E_{\textbf{k},TM}^{x}\left(x,y\right)&=&\frac{\texttt{i} k_{z}k_{m}}{k_{m}^{2}+k_{n}^{2}} \nonumber \\
  &\times& B_{mn}\cos{\left(k_{m}x\right)}\sin{\left(k_{n}y\right)},
\end{eqnarray}
\begin{eqnarray}\label{9}
  E_{\textbf{k},TM}^{y}\left(x,y\right)&=&\frac{\texttt{i} k_{z}k_{n}}{k_{m}^{2}+k_{n}^{2}} \nonumber \\
  &\times& B_{mn}\sin{\left(k_{m}x\right)}\cos{\left(k_{n}y\right)},
\end{eqnarray}
\begin{equation}\label{10}
  E_{\textbf{k},TM}^{z}\left(x,y\right)=B_{mn}\sin{\left(k_{m}x\right)}\sin{\left(k_{n}y\right)}.
\end{equation}
Here $k_m=m\pi/a$, $k_n=n\pi/b$, $k=\sqrt{k_{m}^{2}+k_{n}^{2}+k_{z}^{2}}=\omega_{k}/c$. The indexes $m$ and $n$ are positive integers for TM modes, and for TE modes $m,n=0,1,2,...$, herewith both indexes cannot be zero together. $B_{mn}$ is the normalization constant, which can be obtained on the
basis of the standard form of the field Hamiltonian. Reference point is chosen at one of the corners of the cross section, so the space into a waveguide corresponds to the positive values of the coordinates $x$ and $y$, see Fig. 1.

\begin{figure}\center
	\includegraphics[width=10cm]{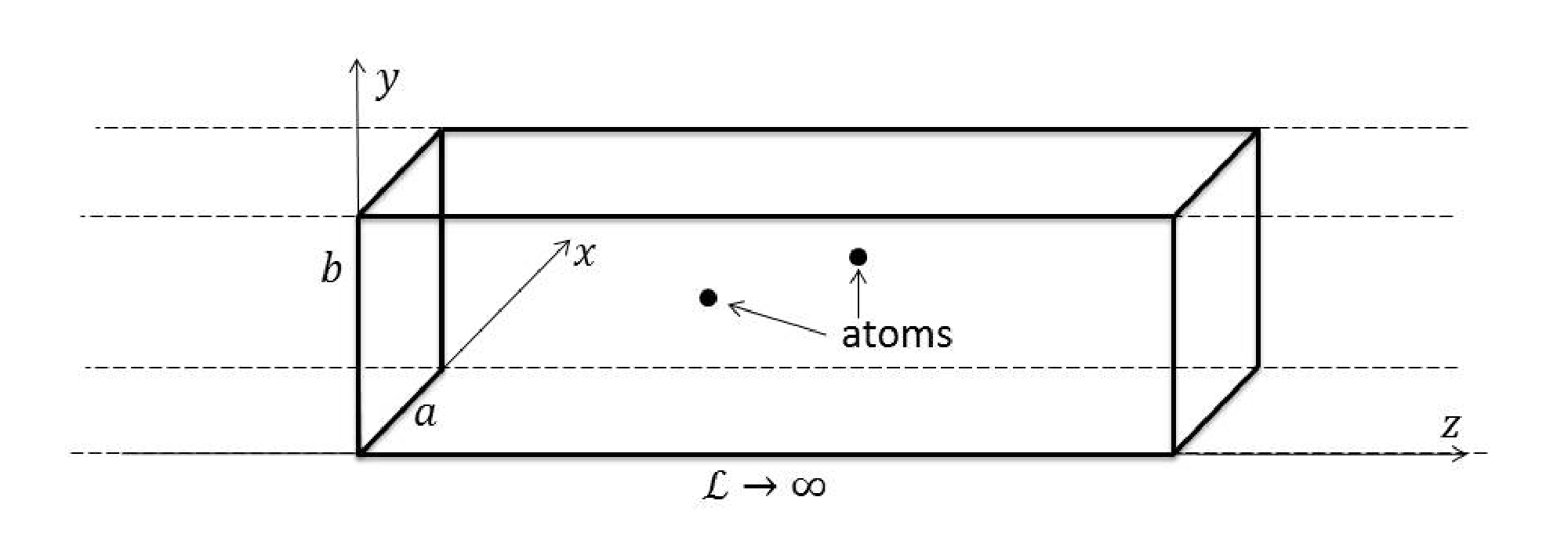}
	\caption{\label{fig:sk}
			Sketch of the waveguide and the atoms inside it.}\label{sk}
\end{figure}

Formally solving the Schrodinger equation for the system "atoms+field" and restricting
ourselves by the states containing no more than one photon (i.e. neglecting nonlinear effects), one obtains a system of equations for the amplitudes $b_e$ of one-fold atomic excited states with the coupling
between atoms described by the so-called "Green's matrix" \cite{SKH2011}. It is essentially built up of Green's functions of
Maxwell equations, describing the propagation of light in a waveguide
from one atom to another. This $3N\times3N$ matrix plays a key role in the theory, describing both single-atom effects and the radiative transfer between different atoms.

According to the general quantum microscopic approach essentially based on the coupled-dipole model, the Green's matrix $G_{ee'}(\omega)$ is given as follows:
\begin{eqnarray}\label{16}
  G_{ee'}(\omega)&&=-\frac{2}{\gamma_0}\biggl\{\sum_{g}V_{e;g}V_{g;e'}\zeta(\hbar\omega-E_{g})+ \nonumber
  \\
  &&\sum_{ee}V_{e;ee}V_{ee;e'}\zeta(\hbar\omega-E_{ee})\biggl\}.
\end{eqnarray}
This equation includes matrix elements of the operator $\widehat{V}$ of the interaction between atoms and electromagnetic field, $\zeta(x)$ is a singular function which is determined
by the relation $\zeta\left(x\right)=\lim\limits_{k\to\infty}{(1-\exp(ikx))/x}$. To calculate the Green's matrix, we should perform a summation over resonant single-photon states "$g$" as well as over non-resonant states with two excited atoms and one photon "$ee$" (as greater length, see \cite{SKH2011}). Actually, this approach allows one to describe from a single position both monatomic dynamics and cooperative effects caused by interatomic dipole-dipole interaction. The main idea of this approach was first proposed by Foldy \cite{Foldy}, further it was developed by a number of authors, to name a few \cite{Stephen,Hameka,Bonifacio1,Bonifacio2,Ficek}. This method
was successfully used in our group for the analysis of the optical properties of dense atomic ensembles as well as for studying light scattering from such ensembles \cite{KS_PRA_2011,KS_J_Mod_Opt_2013,KS_PRA_2015,Roof15,Skipetrov_2016,KS_Las_Phys_2017,KS_PRA_2017}. Further it allowed us to describe cooperative effects in atomic ensembles located in a Fabry-Perot cavity \cite{KS2016J,KS2016} and near a conducting surface \cite{KS2018a,KS2018b,KS2018c}.

The calculation of the explicit expressions for the Green's matrix corresponding to a waveguide is provided in the Appendix.

\section{Results and discussion}
\subsection{Single-atom effect}

We first analyze the spontaneous decay dynamics of a single excited atom placed in a waveguide. The character of decay dramatically depends on the transverse sizes of a waveguide $a$ and $b$, because these sizes determine the cutoff frequency. Without any restriction of generality, we assume $a\geq b$. It is known that if the resonant frequency of atomic transition $\omega_{0}$ is less that the cutoff frequency of a waveguide $\omega^{c}$, then single-atom spontaneous decay is totally suppressed for any Zeeman sublevel \cite{Kleppner}. This is explained by the fact that in such a case, no one field mode at the transition frequency can propagate in a waveguide as oscillating wave. Different modes have different cutoff frequencies, and $\omega^{c}$ is determined by the mode, which has a minimal one. In the considered case, it is $\text{TE}_{10}$ mode, so $\omega^{c}=\omega_{10}^{c}=c\pi/a$. When $\omega_{0}>\omega_{10}^{c}$, single-atom spontaneous decay is allowed, but the character of this decay depends on the Zeeman sublevel which was initially populated. The developed theory allows us to consider arbitrary initial condition. From the experimental point of view, this is determined by the technique employed to excite atoms. Generally, an atom can be prepared in a superposition of the ground state and all the Zeeman sublevels of the excited state. For clarity, we first consider the case when at initial time only one Zeeman sublevel $m_{J}=-1$ is populated with 100\% probability. Electromagnetic field is initially in the vacuum state.

Figure 2 shows the population dynamics of all Zeeman sublevels of the atomic excited state $P_{e}(t)=|b_{e}(t)|^{2}$ calculated for given initial condition. The transverse sizes of a waveguide were chosen $a=4$, $b=2$ (hereafter we consider the inverse wavenumber of radiation resonant to the atomic transition $k_{0}^{-1}=c/\omega_{0}$ as a unit of length). In this case, the transition frequency $\omega_0$ significantly exceeds the cutoff frequency of a waveguide, $\omega_0/\omega^{c}=4/\pi\approx1.27$, and it is far from the vicinities of the cut-offs of all the waveguide modes (under the considered assumption of perfectly conducting walls of a waveguide, these vicinities are infinitely narrow). The waveguide with given transverse sizes is single-mode, because the cutoff frequencies of all the modes except $\text{TE}_{10}$ exceed $\omega_{0}$ (see the inset in Fig. 2). Thus, only $\text{TE}_{10}$ mode is responsible for the spontaneous decay. An atom was considered at the axis of a waveguide. In the Fig. 2 we see that excitation probability of the sublevel $m_{J}=-1$ decreases with time from its initial value equal to 1 and asymptotically approaches to 0.25 at large times. Thus, we observe incomplete spontaneous decay. It is obvious that the well-known mechanism of decay suppression which was described in Refs. \cite{Kleppner} and \cite{Lambropoulos}, namely frequency selection, cannot explain the observed effect under considered conditions. Moreover, we see the gradual population of another Zeeman sublevel -- $m_{J}=1$. Its excitation probability grows from the initial value equal to 0, and approaches at large times to the same asymptotic value 0.25. The sublevel $m_{J}=0$ of the excited state does not populate with time.

\begin{figure}\center
	\includegraphics[width=7cm]{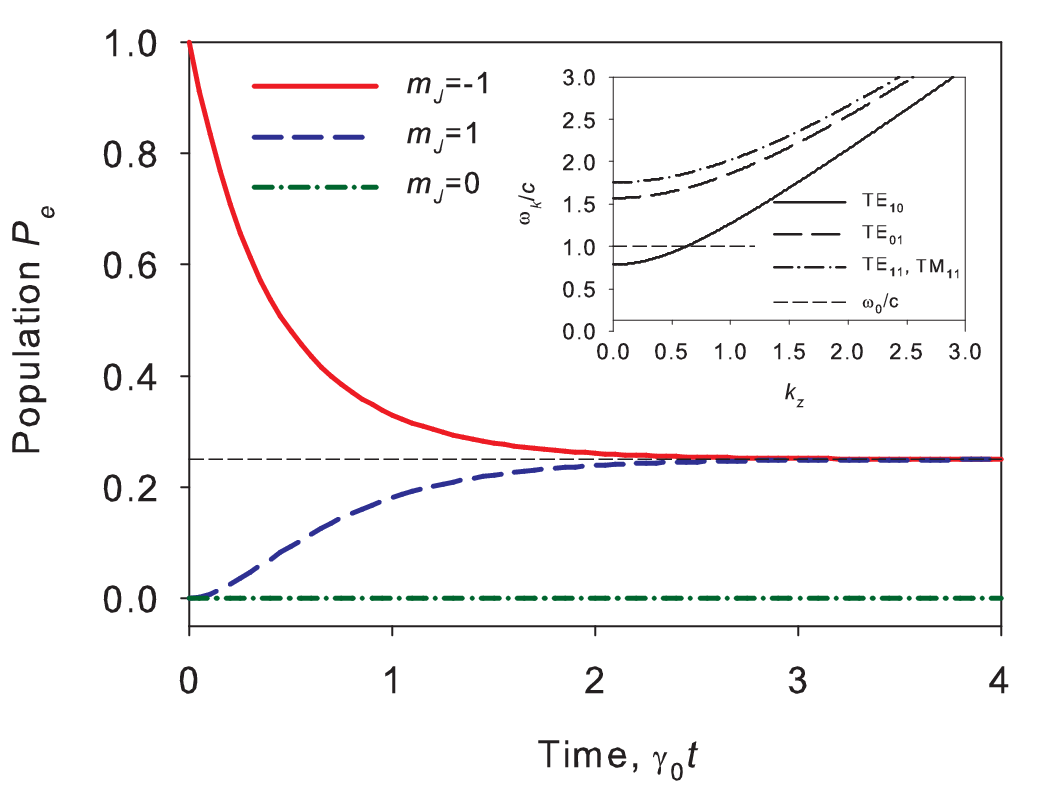}
	\caption{\label{fig:two}
			Population of different Zeeman sublevels of the atomic excited state depending on time. The transverse sizes of a waveguide $a=4$, $b=2$. An atom is located at the axis of a waveguide. At $t=0$ only one sublevel $m_{J}=-1$ is populated.}\label{f2}
\end{figure}

To explain the nature of the dependencies plotted in the Fig. 2, let us take into account that the process of spontaneous decay is caused by the multiple emission and subsequent absorption of virtual photons. The vector of electric field in the mode $\text{TE}_{10}$ has only one nonzero component along $y$ axis that is clear from Eqs. (\ref{5}) -- (\ref{7}). Therefore, upon the atomic transition from the sublevel $m_{J}=-1$ to the ground state, only half of the energy of atomic excitation transfers to field subsystem with $y$-polarized photon. The subsequent absorption of $y$-polarized virtual photon leads to an equiprobable excitation of the sublevels $m_{J}=-1$ and $m_{J}=1$. Thus, as a result of the multiple emission and absorption, a half of the energy of atomic excitation transfers to the electromagnetic field, and remaining half is equally distributed among the sublevels $m_{J}=-1$ and $m_{J}=1$. The sublevel $m_{J}=0$ does not populate because the electric field in $\text{TE}_{10}$ has no $z$-component. We call the described mechanism as "polarization selection". In the considered example, the dipole momentum of the transition from the excited state $m_{J}=-1$ to the ground state has only one circular component $\sigma^{-}$, which can be presented as a superposition of two linear components $x$ and $y$. $y$-component is decaying, while the decay of $x$-component is forbidden because $\text{TE}_{10}$ mode has no corresponding component of the electric field. In fact, the effect of polarization selection occurs upon the decay of such a superposition state, in which some components are decaying while others correspond to the dark state, which is non-decaying due to polarization effects.

Note that the effects described here can be correctly explained and clearly understood only if the polarization properties are taken into account. In the literature, one can find different approaches to the description of the dynamics of a two-level system (atom) embedded into a waveguide. In a number of them, 1D approximation for the photon mode involved in the interaction is used, see for example Ref. \cite{Longo}. In general, the description of polarization effects requires a realistic 3D model for the photon modes. This allows us to describe correctly the atomic dynamics both in a single-mode and in a multimode waveguide.

Since the Green's matrix is only $3\times3$ in the case of a single atom, we are able to obtain the analytical expressions for the quantum amplitudes of the one-fold atomic excites states $b_{e}(t)$ and, consequently, $P_{e}(t)$. For the sublevel $m_{J}=-1$:
\begin{eqnarray}
b_{e}(t)=\frac{\texttt{i}}{2}\left(1+\exp\left(-\frac{\gamma'}{2}t\right)\right), \nonumber
\end{eqnarray}
\begin{eqnarray} P_{e}(t)=\frac{1}{4}\left(1+2\exp\left(-\frac{\gamma'}{2}t\right)+\exp\left(-\gamma't\right)\right). \nonumber
\end{eqnarray}
Thus, the spontaneous decay dynamics is described by biexponential law. For the considered parameters $\gamma'\approx3.8\gamma_{0}$.

For Zeeman sublevel $m_{J}=1$ the dynamics is described as follows:
\begin{eqnarray}
b_{e}(t)=\frac{\texttt{i}}{2}\left(-1+\exp\left(-\frac{\gamma'}{2}t\right)\right), \nonumber
\end{eqnarray}
\begin{eqnarray}
P_{e}(t)=\frac{1}{4}\left(1-2\exp\left(-\frac{\gamma'}{2}t\right)+\exp\left(-\gamma't\right)\right). \nonumber
\end{eqnarray}

Note that the curves shown in Fig. 2 were obtained upon specific initial condition -- when only one sublevel $m_{J}=-1$ was excited at $t=0$. Of course, upon another initial condition the results will differ. Thus, if the sublevel $m_{J}=0$ of the excited state is initially populated, then spontaneous decay is totally suppressed, i.e. $P_{e}(t)\equiv1$ for $m_{J}=0$ and $P_{e}(t)\equiv0$ for $m_{J}=\pm1$. This is explained by the fact that spontaneous decay of given sublevel requires the presence of field modes with nonzero $z$-component that can propagate in a waveguide as oscillating wave. In the case when the sublevel $m_{J}=1$ is initially populated, the results are mirror symmetrical to those shown in Fig. 1 (i.e. the curves for $m_{J}=-1$ and $m_{J}=1$ are swapped).

The rate of incomplete spontaneous decay is determined by the parameter $\gamma'$. It depends on the position of the atom in the plane perpendicular to the axis of a waveguide (there is no dependence on its $z$-position, because all the point along $z$ axis are physically equal in an infinite waveguide). By the analysis of the eigenvalues of Green's matrix, we derived the expression for $\gamma'$:
\begin{equation}\label{11} \gamma'=\frac{6\pi\gamma_{0}}{k_{0}^{2}ab\sqrt{1-\left(\frac{\pi}{k_{0}a}\right)^{2}}}\sin^{2}\left(\frac{\pi x_{1}}{a}\right),
\end{equation}
where $x_{1}$ means $x$ coordinate of the atom. Note that $\gamma'$ does not depend on its $y$ coordinate. The reason of this feature is that electric field in the $\text{TE}_{10}$ mode has no dependence on $y$, that is clear from Eq. (\ref{6}).

To verify the explanation of the observed effect, we changed the frequency of atomic transition. The alteration in $\omega_0$, if it does not cross the cut-offs of the waveguide modes, cannot qualitatively change the picture.

We carried out the calculations of the atomic excitation dynamics in a waveguide with different transverse sizes. Our analysis shows that in the case of a multimode waveguide, the effect of incomplete spontaneous decay disappears. For instance, when $a=b=8$ (in such a waveguide 10 modes at the transition frequency can propagate long distances along the axis: $\text{TE}_{10}$, $\text{TE}_{01}$, $\text{TE}_{20}$, $\text{TE}_{02}$, $\text{TE}_{11}$, $\text{TE}_{12}$, $\text{TE}_{21}$, $\text{TM}_{11}$, $\text{TM}_{12}$, $\text{TM}_{21}$), we observe that upon the spontaneous decay of any Zeeman sublevel $m_{J}$, other sublevels are almost not populated. Herewith, all the energy of initial excitation transfers to the electromagnetic field, and the decay dynamics can be described by a traditional single-exponential law with a good accuracy.

\subsection{Diatomic effect}

Alteration of the spatial structure of modes of electromagnetic field in a waveguide leads not only to the modification of single-atom properties, but also qualitatively changes the character of any electromagnetic interaction between different atoms, in particular, the most pronounced dipole-dipole interaction. This effect also dramatically depends on the ratio between the transition frequency $\omega_0$ and the cutoff frequency of a waveguide $\omega^{c}$. In the case of $\omega_0<\omega^{c}$, photon exchange between atoms is caused by near-field effects, so at long distances it is suppressed \cite{Passante}. This situation is a lot like that taking place in a Fabry-Perot cavity with small separation between the mirrors, when single-atom spontaneous decay of some Zeeman sublevels is suppressed, but near-field energy exchange between different atoms recovers decay dynamics \cite{KS2016J,KS2016}. In the opposite case, $\omega_0>\omega^{c}$, the dipole-dipole interaction is essentially long-range, and the dynamics of a given atom can be significantly affected even by far-distant atoms \cite{waveduide_group2_3}.

Let us consider two atoms in a single-mode waveguide with transverse sizes $a=4$, $b=2$. First atom is located at the point $x_1$, $y_1$, $z_1$; second -- $x_2$, $y_2$, $z_2$. We assume that at initial time only one Zeeman sublevel $m_{J}=-1$ of the first atom is populated. Second atom is in the ground state.

Figure 3 shows the dynamics of the total excited state population $P_{sum}(t)$, calculated as a sum of $P_{e}(t)$ over all the Zeeman sublevels, separately for the first and for the second atom. In order to compare, we show the curve corresponding to single-atom case, when second atom is absent. Fig. 3 demonstrates two main results. The first one is that the effect of incomplete spontaneous decay takes plays in the diatomic problem. The total excited state population of the first atom in the presence of second one asymptotically approaches to 0.5 at large times (we have checked it at any time scale). Second result is that a significant energy exchange between the atoms takes place even when the interatomic distance is very large. For the considered parameters, interatomic separation is seventeen times greater than resonant wavelength. In free space, the dipole-dipole interaction at such distances is negligible. Herewith, in a waveguide, the population of the excited state of second atom reaches 10\%. Accordingly, one can see a significant difference in the dynamics of the first atom for the cases of presence and absence of second one.

\begin{figure}\center

	\includegraphics[width=7cm]{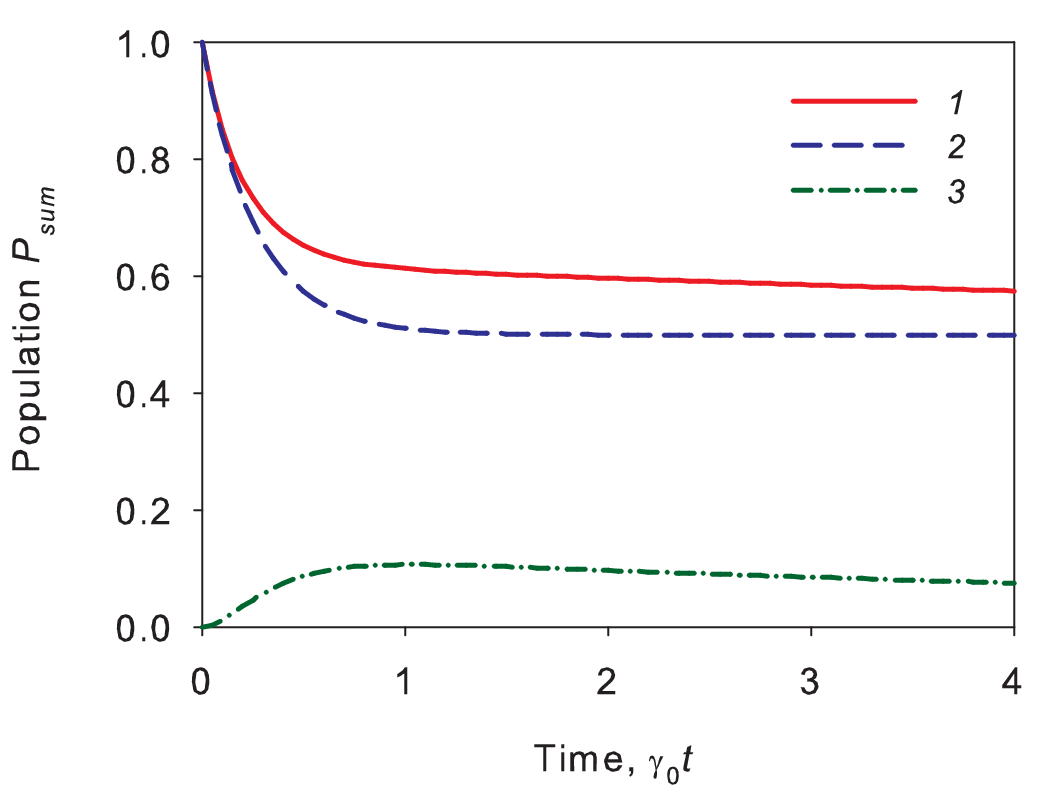}
	\caption{\label{fig:Three}
	Population of the excited state, $a=4$, $b=2$, $x_{1}=x_{2}=a/2$, $y_{1}=y_{2}=b/2$, $z_{2}-z_{1}=107$, \emph{1} -- for the first atom in the presence of second one, \emph{2} -- first atom in the absence of second one, \emph{3} -- second atom.}\label{f3}
\end{figure}

We have studied the maximal population of the excited state of second atom $\max(P_{sum,2})$ depending on the interatomic separation along the axis of a waveguide $\Delta z=z_{2}-z_{1}$, see Fig. 4. We have analyzed the case when both atoms are located at the axis of a waveguide, as well as when the first atom is at the axis and second one at another position. Figure 4 demonstrates a periodical character of the long-range dipole-dipole interaction. In a single-mode waveguide that we considered, the spatial period is $k_{0}T_{s}=\pi/\sqrt{1-\left(\pi/(k_{0}a)\right)^{2}}$, it is 2 times less than the spatial period of $\text{TE}_{10}$ wave at the transition frequency. We noticed an interesting feature, that the maximum of the dependence shown in Fig. 4 is, surprisingly, exactly $1/8$ for the case when both atoms are at $z$ axis and $1/9$ for the second considered case.

\begin{figure}\center
	\includegraphics[width=7cm]{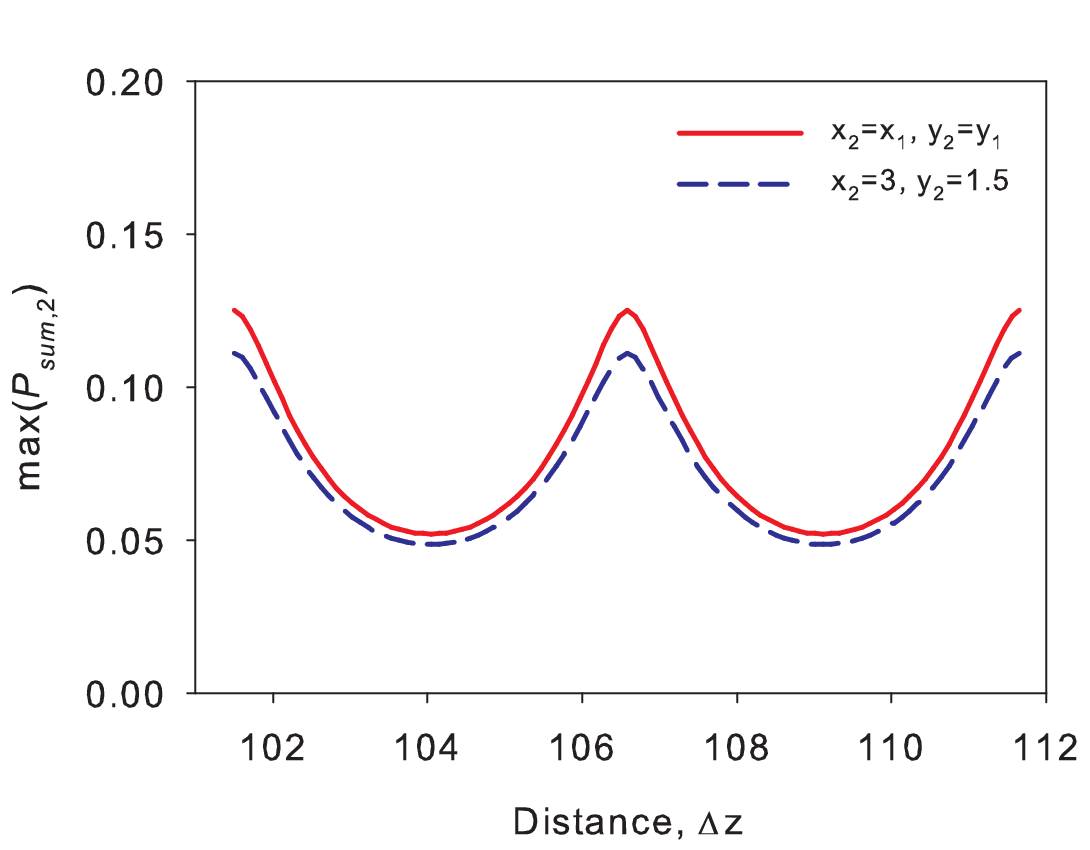}
	\caption{\label{fig:Four}
			Maximal population of the second atom,  $a=4$, $b=2$, $x_{1}=a/2$, $y_{1}=b/2$.}\label{f4}
\end{figure}

In the diatomic case, the Green's matrix has a size of $6\times6$, so obtaining analytical expressions is much more complicated than in the case of a single atom. However, our analysis shows that in the specific case of far-distant atoms in a single-mode waveguide, the Green's matrix has only two nonzero eigenvalues:
\begin{equation}\label{12}
\lambda_{1,2}=r_1+r_3\pm\sqrt{\left(r_1-r_3\right)^2+4r_2^2},
\end{equation}
where $r_1=-\texttt{i}\gamma'/4$, $\gamma'$ is given by Eq. (\ref{11}), $r_3=-\texttt{i}\gamma''/4$, $\gamma''$ is determined by the same equation as $\gamma'$ substituting $x_2$ instead of $x_1$, and
\begin{eqnarray}\label{13}
r_2&=&-\frac{3\texttt{i}\pi\gamma_{0}}{2k_{0}^{2}ab\sqrt{1-\left(\frac{\pi}{k_{0}a}\right)^{2}}}\sin\left(\frac{\pi x_{1}}{a}\right)\sin\left(\frac{\pi x_{2}}{a}\right)\nonumber \\
  &\times&\exp\left(\texttt{i}|z_2-z_1|\sqrt{1-\left(\frac{\pi}{k_{0}a}\right)^{2}}\right).
\end{eqnarray}
Both $\lambda_{1}$ and $\lambda_{2}$ are complex numbers, $\text{Im}(\lambda_{1,2})<0$. The dynamics of the quantum amplitudes of the one-fold atomic excites states is given as follows:
\begin{eqnarray}\label{14}
b_{e_{i,m_{J}}}(t)&=&u_{i,m_{J}}+v_{i,m_{J}}\exp(-\texttt{i}\lambda_{1}t)\nonumber \\
&+&w_{i,m_{J}}\exp(-\texttt{i}\lambda_{2}t).
\end{eqnarray}
To obtain the coefficients $u_{i,m_{J}}$, $v_{i,m_{J}}$ and $w_{i,m_{J}}$, we need to solve the system of linear algebraic equations with full matrix $6\times6$ numerically.

\section{Conclusion}
In conclusion, we considered the dynamics of atomic excitation prepared in a waveguide. We found out the effect of incomplete spontaneous decay -- when the excited state population asymptotically approaches to a nonzero value at large times, under the conditions when the atomic transition frequency is larger than the cutoff frequency of a waveguide and far from the vicinities of the cut-offs. Discovered effect is explained by polarization selection. It has been predicted in a single-mode waveguide with rectangular cross section both for single-atom case and for diatomic case when the long-range dipole-dipole interaction significantly affects the atomic dynamics.

\section{Appendix}
As we assume an infinite length of a waveguide, the sum over the field variables in
Eq. (\ref{16}) should be calculated in the limit of infinite length of the quantization volume along $z$ axis, $L_{q}\to\infty$. This implies summation
over the types of field modes in a waveguide (TE and TM), summation over the transverse indexes $m$ and $n$ as well as the integration over continuous variable $k_z$:
\begin{equation}
\nonumber \sum_{g} \text{or} \sum_{ee}\rightarrow\frac{L_q}{2\pi}\sum_{TE,TM}\sum_{m,n}\int\limits_{-\infty}^{+\infty}dk_{z}.
\end{equation}

To simplify the calculations, it is convenient to perform summation by separate parts:

1) over TE modes with $n=0$,

2) over TE modes with $m=0$,

3) over TE modes with positive integer $m$ and $n$,

4) over TM modes (with positive integer $m$ and $n$).

In accordance with this decomposition, we denote the part of the Green's matrix $G_{ee'}(\omega)$, which is calculated by the sum over the modes of the first group, as $G_{ee'}^{I}(\omega)$; second, $G_{ee'}^{II}(\omega)$; third, $G_{ee'}^{III}(\omega)$; and fourth, $G_{ee'}^{IV}(\omega)$. Applying the so-called polar approximation (i.e.,
neglecting retardation effects), one obtains the following expressions.

\begin{center}
\emph{sum over TE modes with $n=0$}
\end{center}
\begin{eqnarray}\label{20}
G_{ee'}^{I}(\omega_{0})&&\Bigl|_{i=j}=8\texttt{i}\pi \frac{d_{e_{j};g_{j}}^{y}d_{g_{i};e_{i}}^{y}}{\gamma_{0}ab} \nonumber
\\
&&\sum_{m=1}^{\left[\left[\frac{k_{0}a}{\pi}\right]\right]}
\frac{k_{0}^{2}}{\sqrt{k_{0}^{2}-k_{m}^{2}}}\sin^{2}{\left(k_{m}x_{i}\right)},
\end{eqnarray}

\begin{eqnarray}\label{21}
G_{ee'}^{I}(\omega_{0})\Bigl|_{i\neq j}=8\pi \frac{d_{e_{j};g_{j}}^{y}d_{g_{i};e_{i}}^{y}}{\gamma_{0}ab} \nonumber \\
\Biggl\{
\texttt{i}\sum_{m=1}^{\left[\left[\frac{k_{0}a}{\pi}\right]\right]}\sin{\left(k_{m}x_{j}\right)}\sin{\left(k_{m}x_{i}\right)}\nonumber\\
\exp\left(\texttt{i}|z_j-z_i|\sqrt{k_{0}^{2}-k_{m}^{2}}\right)\frac{k_{0}^{2}}{\sqrt{k_{0}^{2}-k_{m}^{2}}}+\nonumber\\
+\sum_{m=\left[\left[\frac{k_{0}a}{\pi}\right]\right]+1}^{+\infty}\sin{\left(k_{m}x_{j}\right)}\sin{\left(k_{m}x_{i}\right)}\nonumber\\
\exp\left(-|z_j-z_i|\sqrt{k_{m}^{2}-k_{0}^{2}}\right)\frac{k_{0}^{2}}{\sqrt{k_{m}^{2}-k_{0}^{2}}}
\Biggl\}.
\end{eqnarray}
Here the index $i$ denotes quantities corresponding
to the atom which transit from excited state to ground one, and the index $j$ is related to the atom which performs reverse transition. Double brackets means the integer part, $\texttt{i}$ means imaginary unit.

\begin{center}
\emph{sum over TE modes with $m=0$}
\end{center}
\begin{eqnarray}\label{24}
G_{ee'}^{II}(\omega_{0})&&\Bigl|_{i=j}=8\texttt{i}\pi \frac{d_{e_{j};g_{j}}^{x}d_{g_{i};e_{i}}^{x}}{\gamma_{0}ab}\nonumber \\
&&\sum_{n=1}^{\left[\left[\frac{k_{0}b}{\pi}\right]\right]}
\frac{k_{0}^{2}}{\sqrt{k_{0}^{2}-k_{n}^{2}}}\sin^{2}{\left(k_{n}y_{i}\right)},
\end{eqnarray}

\begin{eqnarray}\label{25}
G_{ee'}^{II}(\omega_{0})\Bigl|_{i\neq j}=8\pi \frac{d_{e_{j};g_{j}}^{x}d_{g_{i};e_{i}}^{x}}{\gamma_{0}ab}\nonumber \\
\Biggl\{
\texttt{i}\sum_{n=1}^{\left[\left[\frac{k_{0}b}{\pi}\right]\right]}\sin{\left(k_{n}y_{j}\right)}\sin{\left(k_{n}y_{i}\right)}\nonumber \\
\exp\left(\texttt{i}|z_j-z_i|\sqrt{k_{0}^{2}-k_{n}^{2}}\right)\frac{k_{0}^{2}}{\sqrt{k_{0}^{2}-k_{n}^{2}}}+\nonumber \\
+\sum_{n=\left[\left[\frac{k_{0}b}{\pi}\right]\right]+1}^{+\infty}\sin{\left(k_{n}y_{j}\right)}\sin{\left(k_{n}y_{i}\right)}\nonumber \\
\exp\left(-|z_j-z_i|\sqrt{k_{n}^{2}-k_{0}^{2}}\right)\frac{k_{0}^{2}}{\sqrt{k_{n}^{2}-k_{0}^{2}}}
\Biggl\}.
\end{eqnarray}

\begin{center}
\emph{sum over TE modes with positive integer $m$ and $n$}
\end{center}
\begin{eqnarray}\label{26}
G_{ee'}^{III}&&(\omega_{0})\Bigl|_{i=j}=\frac{16\texttt{i}\pi }{\gamma_{0}ab}\nonumber \\
&&\sum_{m,n:\sqrt{k_{m}^{2}+k_{n}^{2}}<k_{0}}
\frac{D_{mn}^{III}}{\sqrt{k_{0}^{2}-k_{m}^{2}-k_{n}^{2}}},
\end{eqnarray}

\begin{eqnarray}\label{27}
G_{ee'}^{III}(\omega_{0})\Bigl|_{i\neq j}=\frac{16\pi}{\gamma_{0}ab}\nonumber \\
\Biggl\{
\texttt{i}\sum_{m,n:\sqrt{k_{m}^{2}+k_{n}^{2}}<k_{0}}\frac{D_{mn}^{III}}{\sqrt{k_{0}^{2}-k_{m}^{2}-k_{n}^{2}}}\nonumber \\
\exp\left(\texttt{i}|z_j-z_i|\sqrt{k_{0}^{2}-k_{m}^{2}-k_{n}^{2}}\right)+\nonumber \\
+\sum_{m,n:\sqrt{k_{m}^{2}+k_{n}^{2}}>k_{0}}\frac{D_{mn}^{III}}{\sqrt{k_{m}^{2}+k_{n}^{2}-k_{0}^{2}}}\nonumber \\
\exp\left(-|z_j-z_i|\sqrt{k_{m}^{2}+k_{n}^{2}-k_{0}^{2}}\right)
\Biggl\},
\end{eqnarray}

\begin{eqnarray}\label{28}
D_{mn}^{III}&=&\frac{k_{0}^{2}}{k_{m}^{2}+k_{n}^{2}}\Bigl[k_{n}d_{e_{j};g_{j}}^{x}\cos{\left(k_{m}x_{j}\right)}\sin{\left(k_{n}y_{j}\right)}-\nonumber \\
&-&k_{m}d_{e_{j};g_{j}}^{y}\sin{\left(k_{m}x_{j}\right)}\cos{\left(k_{n}y_{j}\right)}
\Bigl]\times\nonumber \\
&\times&
\Bigl[k_{n}d_{g_{i};e_{i}}^{x}\cos{\left(k_{m}x_{i}\right)}\sin{\left(k_{n}y_{i}\right)}-\nonumber \\
&-&k_{m}d_{g_{i};e_{i}}^{y}\sin{\left(k_{m}x_{i}\right)}\cos{\left(k_{n}y_{i}\right)}
\Bigl].
\end{eqnarray}

\begin{center}
\emph{sum over TM modes (with positive integer $m$ and $n$)}
\end{center}
\begin{eqnarray}\label{29}
&G&_{ee'}^{IV}(\omega_{0})\Bigl|_{i=j}=\frac{16\texttt{i}\pi }{\gamma_{0}ab}\nonumber \\
&&\sum_{m,n:\sqrt{k_{m}^{2}+k_{n}^{2}}<k_{0}}
D_{mn}^{(IV)1}\sqrt{k_{0}^{2}-k_{m}^{2}-k_{n}^{2}}+\nonumber \\
&+&\frac{D_{mn}^{(IV)3}}{\sqrt{k_{0}^{2}-k_{m}^{2}-k_{n}^{2}}},
\end{eqnarray}

\begin{eqnarray}\label{30}
G_{ee'}^{IV}(\omega_{0})\Bigl|_{i\neq j}=-\frac{16\pi}{\gamma_{0}ab}
\Biggl\{
\left(-\texttt{i}\right) \sum_{m,n:\sqrt{k_{m}^{2}+k_{n}^{2}}<k_{0}} \exp\left(\texttt{i}|z_j-z_i|\sqrt{k_{0}^{2}-k_{m}^{2}-k_{n}^{2}}\right)\nonumber \\
\Biggl[D_{mn}^{(IV)1}\sqrt{k_{0}^{2}-k_{m}^{2}-k_{n}^{2}}
+\texttt{i} D_{mn}^{(IV)2}sign(z_j-z_i) +\frac{D_{mn}^{(IV)3}}{\sqrt{k_{0}^{2}-k_{m}^{2}-k_{n}^{2}}}\Biggl]+\nonumber \\
+\sum_{m,n:\sqrt{k_{m}^{2}+k_{n}^{2}}>k_{0}}
\exp\left(-|z_j-z_i|\sqrt{k_{m}^{2}+k_{n}^{2}-k_{0}^{2}}\right)\nonumber \\
\Biggl[D_{mn}^{(IV)1}\sqrt{k_{m}^{2}+k_{n}^{2}-k_{0}^{2}}
+D_{mn}^{(IV)2}sign(z_j-z_i) -\frac{D_{mn}^{(IV)3}}{\sqrt{k_{m}^{2}+k_{n}^{2}-k_{0}^{2}}}\Biggl]
\Biggl\},
\end{eqnarray}

\begin{eqnarray}\label{31}
D_{mn}^{(IV)1}&=&\frac{1}{k_{m}^{2}+k_{n}^{2}}\times\nonumber \\
&\times&\Bigl[k_{m}d_{e_{j};g_{j}}^{x}\cos{\left(k_{m}x_{j}\right)}\sin{\left(k_{n}y_{j}\right)}+\nonumber \\
&+&k_{n}d_{e_{j};g_{j}}^{y}\sin{\left(k_{m}x_{j}\right)}\cos{\left(k_{n}y_{j}\right)}
\Bigl]\times\nonumber \\
&\times&
\Bigl[k_{m}d_{g_{i};e_{i}}^{x}\cos{\left(k_{m}x_{i}\right)}\sin{\left(k_{n}y_{i}\right)}+\nonumber \\
&+&k_{n}d_{g_{i};e_{i}}^{y}\sin{\left(k_{m}x_{i}\right)}\cos{\left(k_{n}y_{i}\right)}
\Bigl],
\end{eqnarray}

\begin{eqnarray}\label{32}
D_{mn}^{(IV)2}&=&d_{g_{i};e_{i}}^{z}\sin{\left(k_{m}x_{i}\right)}\sin{\left(k_{n}y_{i}\right)}\times\nonumber \\
&\times&\Bigl[k_{m}d_{e_{j};g_{j}}^{x}\cos{\left(k_{m}x_{j}\right)}\sin{\left(k_{n}y_{j}\right)}+\nonumber \\
&+&k_{n}d_{e_{j};g_{j}}^{y}\sin{\left(k_{m}x_{j}\right)}\cos{\left(k_{n}y_{j}\right)}
\Bigl]-\nonumber \\
&-&d_{e_{j};g_{j}}^{z}\sin{\left(k_{m}x_{j}\right)}\sin{\left(k_{n}y_{j}\right)}\times\nonumber \\
&\times&
\Bigl[k_{m}d_{g_{i};e_{i}}^{x}\cos{\left(k_{m}x_{i}\right)}\sin{\left(k_{n}y_{i}\right)}+\nonumber \\
&+&k_{n}d_{g_{i};e_{i}}^{y}\sin{\left(k_{m}x_{i}\right)}\cos{\left(k_{n}y_{i}\right)}
\Bigl],
\end{eqnarray}

\begin{eqnarray}\label{33}
D_{mn}^{(IV)3}&=&d_{e_{j};g_{j}}^{z}d_{g_{i};e_{i}}^{z}\left(k_{m}^{2}+k_{n}^{2}\right)\times\nonumber \\
&\times&\sin{\left(k_{m}x_{j}\right)}\sin{\left(k_{n}y_{j}\right)}\times\nonumber \\
&\times&\sin{\left(k_{m}x_{i}\right)}\sin{\left(k_{n}y_{i}\right)}.
\end{eqnarray}

\section*{Acknowledgments}
This work was supported by the Russian Science Foundation (Grant No. 17-12-01085). A.S.K. appreciates financial support from the Foundation for the Advancement of Theoretical Physics and Mathematics ''BASIS'' and Russian President Grant for Young Candidates of Sciences
(project MK-1452.2020.2).

\end{document}